\documentstyle[twoside]{article}
 
\catcode`\@=11
\long\def\@makefntext#1{
\protect\noindent \hbox to 3.2pt {\hskip-.9pt
$^{{\eightrm\@thefnmark}}$\hfil}#1\hfill}    %CAN BE USED

\def\@makefnmark{\hbox to 0pt{$^{\@thefnmark}$\hss}}  %ORIGINAL
 
\def\ps@myheadings{\let\@mkboth\@gobbletwo
\def\@oddhead{\hbox{}
\rightmark\hfil\eightrm\thepage}
\def\@oddfoot{}\def\@evenhead{\eightrm\thepage\hfil
\leftmark\hbox{}}\def\@evenfoot{}
\def\sectionmark##1{}\def\subsectionmark##1{}}
 
\oddsidemargin=\evensidemargin
\newcounter{sectionc}\newcounter{subsectionc}\newcounter{subsubsectionc}
\renewcommand{\section}[1] {\vspace{12pt}\addtocounter{sectionc}{1}
\setcounter{subsectionc}{0}\setcounter{subsubsectionc}{0}\noindent
   {\tenbf\thesectionc. #1}\par\vspace{5pt}}
\renewcommand{\subsection}[1] {\vspace{12pt}\addtocounter{subsectionc}{1}
   \setcounter{subsubsectionc}{0}\noindent
   {\bf\thesectionc.\thesubsectionc. {\kern1pt \bfit #1}}\par\vspace{5pt}}
\renewcommand{\subsubsection}[1] {\vspace{12pt}\addtocounter{subsubsectionc}{1}
   \noindent{\tenrm\thesectionc.\thesubsectionc.\thesubsubsectionc.
   {\kern1pt \tenit #1}}\par\vspace{5pt}}

\newcommand{\textlineskip}{\baselineskip=13pt}
\newcommand{\smalllineskip}{\baselineskip=10pt}
 
\def\eightcirc{
\begin{picture}(0,0)
\put(4.4,1.8){\circle{6.5}}
\end{picture}}
\def\eightcopyright{\eightcirc\kern2.7pt\hbox{\eightrm c}}

\def\abstracts#1{{
   \centering{\begin{minipage}{4.5in}\baselineskip=10pt\footnotesize
   \parindent=0pt #1\par
   \end{minipage}}\par}}
 
\renewenvironment{thebibliography}[1]
   {\frenchspacing
    \ninerm\baselineskip=11pt
    \begin{list}{\arabic{enumi}.}
   {\usecounter{enumi}\setlength{\parsep}{0pt}
    \setlength{\leftmargin 12.7pt}{\rightmargin 0pt} %FOR 1--9 ITEMS
    \setlength{\itemsep}{0pt} \settowidth
   {\labelwidth}{#1.}\sloppy}}{\end{list}}
 
\newcounter{itemlistc}
\newcounter{romanlistc}
\newcounter{alphlistc}
\newcounter{arabiclistc}

\newcommand{\fcaption}[1]{
        \refstepcounter{figure}
        \setbox\@tempboxa = \hbox{\footnotesize Fig.~\thefigure. #1}
        \ifdim \wd\@tempboxa > 5in
           {\begin{center}
        \parbox{5in}{\footnotesize\smalllineskip Fig.~\thefigure. #1}
            \end{center}}
        \else
             {\begin{center}
             {\footnotesize Fig.~\thefigure. #1}
              \end{center}}
        \fi}
 
\newcommand{\tcaption}[1]{
        \refstepcounter{table}
        \setbox\@tempboxa = \hbox{\footnotesize Table~\thetable. #1}
        \ifdim \wd\@tempboxa > 5in
           {\begin{center}
        \parbox{5in}{\footnotesize\smalllineskip Table~\thetable. #1}
            \end{center}}
        \else
             {\begin{center}
             {\footnotesize Table~\thetable. #1}
              \end{center}}
        \fi}
 
\def\@citex[#1]#2{\if@filesw\immediate\write\@auxout
   {\string\citation{#2}}\fi
\def\@citea{}\@cite{\@for\@citeb:=#2\do
   {\@citea\def\@citea{,}\@ifundefined
   {b@\@citeb}{{\bf ?}\@warning
   {Citation `\@citeb' on page \thepage \space undefined}}
   {\csname b@\@citeb\endcsname}}}{#1}}
 
\newif\if@cghi
\def\cite{\@cghitrue\@ifnextchar [{\@tempswatrue
   \@citex}{\@tempswafalse\@citex[]}}
\def\citelow{\@cghifalse\@ifnextchar [{\@tempswatrue
   \@citex}{\@tempswafalse\@citex[]}}
\def\@cite#1#2{{$\null^{#1}$\if@tempswa\typeout
   {IJCGA warning: optional citation argument
   ignored: `#2'} \fi}}

\def\pmb#1{\setbox0=\hbox{#1}
   \kern-.025em\copy0\kern-\wd0
   \kern.05em\copy0\kern-\wd0
   \kern-.025em\raise.0433em\box0}

\def\fnt#1#2{\footnotetext{\kern-.3em
   {$^{\mbox{\scriptsize #1}}$}{#2}}}
 
\def\fpage#1{\begingroup
\voffset=.3in
\thispagestyle{empty}\begin{table}[b]\centerline{\footnotesize #1}
   \end{table}\endgroup}
 
\font\tenrm=cmr10
\font\tenit=cmti10
\font\tenbf=cmbx10
\font\bfit=cmbxti10 at 10pt
\font\ninerm=cmr9

\font\eightrm=cmr8

\def\qed{\hbox{${\vcenter{\vbox{       %HOLLOW SQUARE
   \hrule height 0.4pt\hbox{\vrule width 0.4pt height 6pt
   \kern5pt\vrule width 0.4pt}\hrule height 0.4pt}}}$}}
 
\def\bar#1{\overline{#1}}
\def\etal{{\it et al.}}
\def\ie{{\it i.e.}}
\def\eg{{\it e.g.}}
\begin{document}
 
\normalsize\textlineskip
\thispagestyle{empty}
\setcounter{page}{1}

\vspace*{0.88truein}
 
\fpage{1}
\centerline{\bf NOVEL QCD EFFECTS IN THE PRODUCTION AND}
\vspace*{0.035truein}
\centerline{\bf DECAY OF QUARKONIUM\footnote{
Work supported by the Department of Energy, contract number
DE--AC03--76SF00515.}}
\vspace*{0.37truein}
\centerline{\footnotesize STANLEY J. BRODSKY}
\vspace*{0.015truein}
\centerline{\footnotesize\it Stanford Linear Accelerator Center}
\baselineskip=10pt
\centerline{\footnotesize\it Stanford University, Stanford, California
94309, USA}
 
\vspace*{0.21truein}
\abstracts{There are many outstanding discrepancies comparing
the predictions of perturbative QCD and measurements of the
rate of production and decay of heavy quark systems. The problems
include the $J/\psi \to \rho \pi$ puzzle, leading charmed particle
effects, the anomalous behavior of the heavy quark sea components of 
structure functions, anomalous nuclear target effects,  and the large
rates observed for single and double quarkonium production at large
$x_F$ and large $p_T.$  I argue that these anomalies may be associated
with nonperturbative effects in the higher Fock structure of hadron
wavefunctions.}
 
\vspace*{1pt}\textlineskip
\section{Introduction}
\vspace*{-0.5pt}
\noindent
Heavy quarks act as classical, nonrelativistic color sources, and thus
they  play an invaluable, simplifying role in illuminating basic
features of QCD, such as the nature of hadronic binding and the
mechanisms underlying production dynamics.  In this talk I will review
a number of heavy-quark strong interaction processes which test novel
and subtle features of the theory.  I also emphasize a number of heavy
quark topics in which experiment and theory are apparently discrepant.
 
\vspace*{1pt}\textlineskip
\section{Quarkonium and the Determination of the QCD Coupling}
\vspace*{-0.5pt}
\noindent 
One of the most important recent developments in the analysis of
nonperturbative QCD has been the precise computation of the $c\bar c$
and $b\bar b$ bound state spectra from lattice gauge theory\cite{ref1}.
In this approach, the exact theory is systematically approximated by an
effective Hamiltonian of non-relativistic heavy quarks interacting with
the full gauge field.  The lattice simulation of the $J/\psi$ and
$\Upsilon$ spectrum leads to remarkably precise constraints on the QCD
coupling:
\begin{equation}
\alpha_V(8.2\ GeV) = 0.1945(30)
\end{equation}
from the $\Upsilon$ spectrum, and
\begin{equation}
\alpha_V(8.2\ GeV) = 0.1940(67)
\end{equation}
from the $J/\psi$ spectrum.
Here $\alpha_V$ is the effective coupling defined from the potential for
the interaction of two heavy test charges:
\begin{equation}
V(Q^2) = -4\pi\, C_F\, \left(\frac{\alpha_V(Q^2)}{Q^2}\right) \ .
\end{equation}
 
The coupling $\alpha_V$ satisfies the usual QCD renormalization group
equation, where the first two perturbative coefficients $\beta_0$ and
$\beta_1$ of the $\beta$-function are universal. The running coupling
$\alpha_V(Q^2)$  can in turn be used as the basic input to predict
other PQCD observables. There are no scale ambiguities when using this
coupling since by definition $\alpha_V$ sums all vacuum polarization
insertions\cite{BLM}. Thus this procedure relates observable to
observable, eliminating scheme and scale ambiguities.  For example,
$\alpha_V(\ell^2)$ enters directly into the calculation of the hard
scattering amplitude $T_H$ which controls exclusive processes at large
momentum transfer, where $\ell^2$ is the four-momentum squared carried
by the gluons. Another important application is the production of heavy
quark systems near threshold\cite{Kuhn}. The cross section for the
exclusive production $e^+e^-\rightarrow Q\bar Q$ at low $Q$-$\bar Q$
relative velocities is given by the Born rate multiplied by a
Sommerfeld factor
\begin{equation}
S(x) = \frac{x}{1-e^{-x}}
\end{equation}
where
\begin{equation}
x = \pi \, C_F \alpha_V (\vec p^2_{\rm rel})
\end{equation}
is evaluated at the relative momentum $\vec p{}^2_{\rm ref} = m^2_Q
v^2_{\rm rel}$ of the heavy quarks.  The PQCD Sommerfeld factor also
leads to a distinctive angular distribution for the $e^+e^- \rightarrow
c\bar c$ and $b\bar b$ systems which should be reflected in the
$\theta_{\rm cm}$ dependence of the corresponding exclusive
channels\cite{Kuhn}.

One can also use the BLM procedure\cite{BLM} to derive a ``commensurate
scale relation"\cite{CSR,ref1} connecting the effective charge  $\alpha_V$ to
the standard 
$\overline {MS}$ coupling:
\begin{equation}
\alpha_{\overline {MS}}(Q) = \alpha_V(e^{5/6}Q)[1 + (2/\pi) \alpha_V +
(0.96 + C n_F) \alpha_V^2+ \cdots]
\end{equation}
The NNLO term has recently been calculated by L\"uscher and
Weisz\cite{luwe}.
Thus one can use the quarkonium spectrum to predict the $\overline
{MS}$ coupling at any scale; \eg, one finds $\alpha_{\overline
{MS}}(M_Z)= 0.117(2)$,\cite{ref1} assuming that the uncalculated NLO $n_F$
coefficient is zero. This equation also provides an analytic extension
of the $\overline {MS}$ coupling to an effective charge with a quark
mass-dependent $\beta$-function\cite{BGM}.
 
\vspace*{1pt}\textlineskip
\section{Anomalies in Charm Hadroproduction}
\vspace*{-0.5pt}
\noindent
Although most predictions of PQCD are in reasonable agreement with
experiment, a number of processes involving charm hadroproduction
appear to be glaring \hbox{exceptions.}  The following is a partial
list of the empirical anomalies together with a brief discussion of
their possible cures:
 
{\it $J/\psi$ production at large $p_T$.}\ 
The cross section for the production of $J/\psi$'s at high transverse
momentum at the Tevatron is a factor $\sim$ 30 above PQCD predictions
based on gluon fragmentation to a color-singlet $c\bar c$
$J/\psi$.\cite{CDF}  The production cross sections for other heavy
quarkonium states also show similar anomalies. As discussed by
Rothstein\cite{ref3} at this meeting, a possible cure is the hypothesis
that the gluon fragmentation proceeds through $c\bar c g$ and $c\bar c
g g$ Fock components in the charmonium light-cone wavefunction in which
the charmed pair is in a color-octet configuration. As noted by
Magnago\cite{ref4},  one cannot assume that the fragmentation function
is a $\delta$-function at $z=1$,  because $g \rightarrow J/\psi\, gg$
fragmentation is softened by the presence of spectator gluons with
finite momenta in the quarkonium rest frame. The magnitude of the
required color-octet matrix elements thus must be somewhat larger than
usually assumed and may be in conflict with other constraints, such as
$J/\psi$ photoproduction\cite{MSchmidt}.

An alternative and apparently phenomenologically successful hypotheses
is the ``color evaporation model"\cite{ref5}. This is a duality
approach, where one simply computes the probability in leading twist
QCD for the production of charmed pairs with invariant mass below the
open charm threshold. It is postulated that the usual constraints of
color conservation for hard processes and fragmentation can be ignored;
\ie, it is assumed that  color is restored via soft interactions with
the spectator partons. An important question is whether this approach
is consistent with the hard scattering factorization theorem, and
whether the leading order rate for Drell--Yan leptons pairs still
requires in this scheme the usual $1/N_C$ factor implied by color
conservation.

{\it The charm structure functions $c(x,Q^2)$.}\ 
The charm structure function of the proton measured by the EMC
collaboration\cite{emc} is some 30 times larger at $x_{Bj}=0.47$,
$Q^2=75\ GeV^2$ than that predicted on the basis of photon-gluon fusion
$\gamma^*g\rightarrow c\bar c$.  This remarkable anomaly needs to be
confirmed by another large $x_{Bj}$ charm and bottom structure function
measurements.

The excess charm signal observed by EMC can be explained by the
``intrinsic charm'' hypotheses\cite{intc} where the charm sea is
derived from $c \bar c$ contributions to the proton light-cone
wavefunction\cite{Bro95} beyond gluon splitting.  Any Feynman diagram
in which the $c\bar c$ is multiply-connected to the valence quarks of
the proton produces a source of intrinsic charm (IC) in the hadron
wavefunction, in distinction to gluon-splitting $g\rightarrow c\bar c$
``extrinsic charm'' diagrams in which the charm quarks are constituents
of the gluon rather than the proton itself.  A crucial feature of the
IC contribution is the fact that the LC wavefunction $c\bar c\, uud$ is
maximal when the five quarks have minimal invariant mass and are thus
at minimal relative rapidity.  The heavy quarks thus tend to be
produced with the largest momentum fractions, thus accounting for the
EMC anomaly.  A recent re-analysis of the EMC data by Vogt, Harris, and
Smith\cite{ref6}, indicates that the probability of IC in the nucleon
is of order $0.6 \pm 0.3\%$. Thus, in general, one must distinguish two
distinct types of quark and gluon contributions to the nucleon sea
measured in deep inelastic lepton-nucleon scattering: ``extrinsic" and
``intrinsic"\cite{intc}. The extrinsic sea quarks and gluons are
created as part of the lepton-scattering interaction and thus exist
over a very short time $\Delta \tau \sim 1/Q$. These factorizable
contributions can be systematically derived from the QCD hard
bremsstrahlung and pair-production (gluon-splitting) subprocesses
characteristic of leading twist perturbative QCD evolution. In
contrast, the intrinsic sea quarks and gluons are multi-connected to
the valence quarks and exist over a relatively long lifetime within the
nucleon bound state. Thus the intrinsic $q \bar q$ pairs can arrange
themselves together with the valence quarks of the target nucleon into
the most energetically-favored meson-baryon fluctuations. The
enhancement of the heavy quark sea at large momentum fractions where
the relative rapidities between the valence and heavy quarks are
minimized has also been observed in solutions of QCD(1+1) with two
flavors using the discretized light-cone quantization
method\cite{Hornbostel}.
 
{\it Asymmetric sea.}\ 
In conventional studies of the ``sea'' quark distributions, it is
usually assumed that, aside from the effects due to antisymmetrization
with valence quarks, the quark and antiquark sea contributions have the
same momentum and helicity distributions. However, the ansatz of
identical quark and anti-quark sea contributions has never been
justified, either theoretically or empirically. Obviously the sea
distributions which arise directly from gluon splitting in leading
twist are necessarily CP-invariant; \ie,\ they are symmetric under
quark and antiquark interchange. However, the initial distributions
which provide the boundary conditions for QCD evolution need not be
symmetric since the nucleon state is itself  not CP-invariant. Only the
global quantum numbers of the nucleon must be conserved. The intrinsic
sources of strange (and charm) quarks reflect the wavefunction
structure of the bound state itself; accordingly, such distributions
would not be expected to be CP symmetric\cite{BrodskyMa}. Thus the
strange/anti-strange asymmetry of nucleon structure  functions provides
a direct window into the  quantum bound-state structure of hadronic
wavefunctions.
 
It is possible to consider the nucleon wavefunction at low resolution
as a fluctuating system coupling  to intermediate hadronic Fock states
such as non-interacting meson-baryon pairs. The most important
fluctuations are those closest to the energy shell with minimal
invariant mass.  For example, the coupling of a proton to a virtual
$K^+ \Lambda$ pair provides a specific source of intrinsic strange
quarks and antiquarks in the proton.   Since the $s$ and $\bar s$
quarks appear in different configurations in the lowest-lying hadronic
pair states, their helicity and momentum distributions are distinct.
Recently Ma and I\cite{BrodskyMa} and I have investigated the quark
and antiquark asymmetry in the nucleon sea which is implied by a
light-cone meson-baryon fluctuation model of intrinsic $q\bar q$ pairs.
Such fluctuations are necessarily part of any  quantum-mechanical
description of the hadronic bound state in QCD and have also been
incorporated into the cloudy bag model \cite{Sig87} and Skyrme
solutions to chiral theories\cite{Bur92}. We have utilized a
boost-invariant light-cone Fock state description of the hadron
wavefunction which emphasizes multi-parton configurations  of minimal
invariant mass. We find that such fluctuations predict a striking sea
quark and antiquark asymmetry in the corresponding momentum and
helicity distributions in the nucleon structure functions.  In
particular, the strange and anti-strange distributions in the nucleon
generally have completely different momentum and spin characteristics.
For example, the model predicts that the intrinsic $d$ and $s$ quarks
in the proton sea are negatively polarized, whereas the intrinsic $\bar
d$ and $\bar s$ antiquarks provide zero contributions to the proton
spin. We also predict that the intrinsic charm and anticharm helicity
and momentum distributions are not strictly identical. The above
picture of quark and antiquark asymmetry in the momentum and helicity
distributions of the nucleon sea quarks has support from a number of
experimental observations, and we suggest processes to test and measure
this quark and antiquark asymmetry in the nucleon sea.
 
More recently, Ma and I\cite{BMnew} have noted that the hadronic jet
fragmentation of the $s$ and $c$ quarks in electron-positron ($e^+e^-$)
annihilation at the $Z^0$-boson resonance also  provides a laboratory
for testing the quark/antiquark asymmetries of the nucleon sea.
Crossing symmetry implies that the asymmetries of the $s \bar s$ and $c
\bar c$ pairs of the nucleon sea will be reflected in the
nucleon/anti-nucleon asymmetries from the hadronic jet fragmentation of
$s$  and $c$ quarks into nucleons and anti nucleons. For example, if
one has a pure sample of tagged $s$ jets, then one can look for the
difference of $D_{p/s}(z)-D_{\bar p/s}(z)$  at large $z$. Here
$D_{h/q}(z)$ is the fragmentation function representing the probability
for the quark $q$ splitting into the hadron $h$ and $z$ is the momentum
fraction carried by the fragmented hadron from the quark jet.
 
{\it Anomalous large $x_F$ $J/\psi$ production.}\  
The CERN experiment NA3\cite{ref7} has reported a number of
experimental analyses in the hadroproduction of $J/\psi$'s.  The most
dramatic feature of the NA3 data is the longitudinal momentum
distribution of {\it pairs} of $J/\psi$'s as determined from
$pA\rightarrow \mu^+\mu^-\mu^+\mu^-X$ and $\pi A\rightarrow
\mu^+\mu^-\mu^+\mu^-X$ events.  The NA3 data shows that the $X^{TOT}_F$
distribution peaks at the highest $X^{TOT}_F$ bins, where $X^{TOT}_F$
is the sum of the two $J/\psi$ momentum fractions. The observed
distributions, although sparse, are in strong contradiction to the
prediction based on PQCD fusion processes which peak at small
$X^{TOT}_F$.  The NA3 distributions can be reproduced within the
intrinsic charm model, assuming that the $J/\psi$ pair events originate
from the diffractive excitation of $|uud\, c\bar c\, c\bar c\rangle$
intrinsic charm Fock state in the proton\cite{intc,ISR}.  In such
configurations, the four charmed quarks tend to carry almost all of the
momentum of the proton. Only a small momentum transfer to the target is
necessary to put the multi-charm state on-shell in a high energy
collision.  Vogt and I\cite{ref16} have also presented predictions for
$J/\psi$-$\Upsilon$ and $\Upsilon$-$\Upsilon$ pairs based on intrinsic
bottom and charm higher Fock states.
 
{\it Anomalous nuclear dependence of $J/\psi$ production.}\ 
The NA3 experiment\cite{ref7} also reports an anomalous change in the
nuclear-number dependence of the $\pi A \rightarrow J/\psi\, X$ and $pA
\rightarrow J/\psi\, X$ cross sections as $x_F$ varies from the central
to forward fragmentation regions.  The nuclear dependence $A^\alpha$ is
found to be  the ``diffractive-like'' at high $x_F$ with $\alpha\simeq
0.71$ for proton beams and $\alpha \simeq 0.77$ for pion beams, which
is characteristic of production on the front surface of the nucleus. 
This observed $A$-dependence is much stronger than expected from the
shadowing of fusion processes or $J/\psi$ absorption and, in any case,
is incompatible with PQCD factorization since the $A^\alpha$ dependence
is a function of $x_F$ rather than the target parton momentum
fraction\cite{hvs}. The observed nuclear dependence is naturally
explained by the postulate that hadroproduction of $J/\psi$'s at large
$x_F$ originates from the diffractive excitation of the IC Fock states
in the projectile\cite{BHMT}. Since the interaction on the target is
soft, the predicted $A$-dependence is characteristic of conventional
hadron-nucleus interactions.  The strong $A$-dependence of the $J/\psi$
hadroproduction cross section at large $x_F$ is also consistent with
recent Fermilab E789 data\cite{E789}.

An important test of the IC picture can be carried out at HERA in low
momentum transfer electron-proton collisions\cite{BHMT}. If the IC
description is correct, then  quarkonia as well as open charm will be
produced at high longitudinal momentum fractions in the proton
fragmentation region.  Since the electron can strike a valence quark,
the kinematics of the produced charm states will be largely insensitive
to the magnitude of the momentum transfer $Q^2.$
 
{\it Polarization of the $J/\psi$.}\
The Chicago-Iowa-Princeton collaboration\cite{Biino} has measured the
polarization of the $J/\psi$ in $\pi N \rightarrow J/\psi\, X$
interactions.  The results are rather remarkable.  The $J/\psi$ is
found to be unpolarized for almost the entire range of $x_F$, except
for the largest bin at $x_F \cong 0.95$ where the polarization changes
to strongly longitudinal.  Neither the predictions of the color-singlet
model\cite{ref10} nor the color-octet model\cite{ref11} can account for
the absence of polarization at moderate $x_F$.  The strong longitudinal
polarization at $x_F\rightarrow 1$ is similar to the pattern observed in
the Drell-Yan reactions $\pi N \rightarrow \mu^+\mu^-X$
\cite{Biino}, which can be readily explained in terms of higher twist
subprocesses such as $(q\bar q)+q\rightarrow \gamma^*\mu^+\mu^-q$
\cite{ref12}.
 
{\it Open charm hadroproduction.}\ 
One of the most controversial areas of charm hadroproduction is the
data for leading charmed and bottom baryon production at large $x_F$.
Several experiments at the ISR\cite{ISR} have reported prominent
signals for $p p \rightarrow \Lambda_c X$. Similar signals were also
reported by the BIS-2 group at Serpukhov\cite{BIS}.  The ISR group of
Basile \etal\ observed $\Lambda_b$ production at large $x_F$, measuring
two decay channels\cite{Basile}, events which are reported by the
Particle Data Group\cite{PDG}. However, these signals appear to imply
very large integrated total cross sections for charm and bottom
hadroproduction if one extrapolates the large $x_F$ data to all $x_F$
using the standard $(1-x_F)^n$ form with $n \geq 0$. However, this
assumption may be incorrect. An interesting possibility is that the
production cross sections $\frac{d\sigma}{dx_F} \, (p p \rightarrow
\Lambda_c X,\Lambda_b X)$ may not be monotonically falling in $x_F$ but
instead have a peaked structure of large $x_F$ reflecting the
coalescence of intrinsic charm or bottom with the valence quark.  The
PYTHIA string acceleration mechanism would also produce such a peak
from coalescence of the charm quarks with the valence quarks of the
projectile.  It should be noted that cross sections such as $(\Xi
N\rightarrow \Omega X)$ rise dramatically by two orders of magnitude
from $x_F\sim 0$ to $x_F\sim I$.\cite{biagi} The possibility of a structured
distribution in heavy hadron production is now being examined more
carefully by Vogt, Quack, and myself\cite{BQV}.

{\it Polarization correlations at the charm threshold.}\ 
One of the most striking anomalies in QCD is the sudden increase in the
polarization correction $A_{NN}$ observed by Krisch and his
collaborators\cite{Krisch} in large $\theta_{cm}\ p p \rightarrow p p$
elastic scattering at $\sqrt s \sim 5\ GeV$.  Measurements at ANL and
BNL show that the rate for elastic scattering at $\theta_{cm} =
90^\circ$ for incident protons polarized parallel and normal to the
scattering plane rises to  4 times that for anti-parallel scattering. 
This seems all the more remarkable since the net correlation of quark
helicities with the proton helicity is small when it is determined in
inclusive deep inelastic lepton scattering.
 
A most interesting explanation of the Krisch anomaly is that it
reflects the onset of the charm threshold in the intermediate
state\cite{ref13}. At $\sqrt s=5\ GeV$ there is just enough energy to
produce the 8-quark system $uud\, uud\, c\bar c$.  Since the quarks are
all at low relative velocity they can interact strongly.  If there is
an S-wave resonance of the 8 quarks, it will be produced only if the
incident protons have $J=L=S=I$.  In fact, such a state only couples to
the incident proton-proton system  when the incident proton and
polarized parallel are normal to the scattering plane. Guy de Teramond
and I\cite{ref13} have shown that the combination of PQCD quark
interchange plus the $uud\, uud\, c\bar c$ resonance in the $p p
\rightarrow p p$ amplitude can account for the Krisch data provided
that the cross section for charm production $p p \rightarrow c\bar c X$
near threshold is of order of $1\ \mu b$.  A  dramatic rise in the
asymmetry $A_{NN}$ is also measured  in $p p\rightarrow p p$ at large
$\theta_{cm}$ at the strangeness threshold, $\sqrt s\cong 3\ GeV$. This
rise again can be accounted for if $\sigma(p p \rightarrow s\bar s X)
\sim O(1\, mb)$ above the strangeness threshold, which is consistent
with experiment.

{\it Nuclear-bound quarkonium.}\
The possibility of strong interactions between heavy and light quarks
at low relative velocity is undoubtedly a general phenomena in QCD. An
important consequence of such attractive forces is nuclear-bound
quark\-onia\cite{ref14}; \eg, a bound state of a $J/\psi$ to a nucleus.
In fact, Manohar, Luke, and Savage\cite{ref15} have used the operator
product expansion to show that the QCD Van der Waals potential is
sufficiently attractive at low relative velocity in the $s$-wave to
lead to bound states of the $J/\psi$ with heavy nuclei.  It is
conceivable that the binding is strong enough to produce bound states
of heavy quarkonia with light nuclei or even nucleons.
 
An interesting place to search for $J/\psi\, N$ resonances on bound
states is in $B \rightarrow J\psi\, \bar\Lambda$ decays at a
$B$-factory\cite{NavarraQuinnSJB}.  The formation of a $[J/\psi\, p]$
bound state would be signaled by events where the $\bar \Lambda$ is
produced with a nearly monotonic energy.
 
\vspace*{1pt}\textlineskip
\section{The Leading Particle Effect in Charm Hadroproduction and the
Effect of Parton Coalescence\cite{ref16}}
\vspace*{-0.5pt}
\noindent
In leading-twist QCD, the PQCD factorization theorem\cite{fact}
predicts that the fragmentation functions $D_{H/c}(z,Q)$ are
independent of the quantum numbers of both the projectile and target.
However,   strong flavor correlations between the produced particle and
the projectile have been reported in charm production \cite{Agb,Bar}.
For example, in $\pi^- (\overline u d)$ interactions with hadrons or
nuclei, the $D^- (\overline c d)$ $x_f$ distribution is consistently
harder than the $D^+ (c \overline d)$ distribution. The $D^-$ and $D^0
(c \overline u)$ are referred to as ``leading"  charmed mesons while
the $D^+$ and $\overline D^0 (\overline c u)$ are ``non-leading". This
leading behavior  suggests that hadronization at large $x_F$ involves
the coalescence of the produced charm or anticharm quarks with the
spectator quarks of the projectile, just as in exclusive reactions. The
study of this phenomena thus can provide new insights into the coherent
mechanisms controlling the formation of hadrons in QCD.
 
The asymmetry between leading and non-leading charm, has been measured
by the WA82\cite{WA82} and E769\cite{E7692} collaborations. Both
experiments find that the measured asymmetry ${\cal A}(x_f),$
integrated over $p_T$, increases from $\sim 0$ for $x_f$ near zero to
$\sim 0.5$ around $x_f = 0.65$. However, the asymmetry ${\cal
A}(p_T^2),$ integrated over all $x_f$, is found to be small in the
range $0<p_T^2< 10$ GeV$^2.$\cite{E7692}\  These facts are consistent
if the leading charm asymmetry is localized at large $x_F,$ involving
only a small fraction of the total cross section.
 
Perturbative QCD at leading order predicts that $c$ and $\overline c$
quarks are produced with identical distributions.  Next-to-leading
order calculations do give rise to a small charge asymmetry ($\sim$
10\% for $x_f \sim 0.8$) between $\overline c$ and $c$ production due
to $q g$ and $q \overline q$ interference \cite{Been,NDE}.  However,
this charge asymmetry should result in an increase of $D^-$, $\overline
D^0$ production over $D^+$ and $D^0$ at high $x_f$, not a separation
between $D^-$, $D^0$ and $D^+$, $\overline D^0$.
 
How can one explain the origin of leading charm asymmetry within the
context of QCD? It is clear that the produced charm (or anticharm)
quark must combine with a projectile valence quark.  Ordinary jet
fragmentation (\eg, Peterson fragmentation \cite{Pete}) cannot
produce a leading particle asymmetry since it is independent of the
initial state and thus the projectile valence quarks. This is an
essential property of leading-twist factorization. However, one expects
on physical grounds that a charm quark produced by fusion may coalesce
with a comoving spectator valence quark\cite{hwa,bg,bedny,andr,ref16}.
For example, in QED, leptons of opposite charge moving with similar
velocities can be captured into neutral atoms\cite{bgs}.  Since the
capture is significant only at small relative rapidity, $\Delta y$, the
effect on the total rate is higher twist.
 
In leading-twist QCD heavy quarks are produced by the fusion
subprocesses $g g \rightarrow Q \overline Q$ and $q \overline q
\rightarrow Q \overline Q$.   The heavy $Q$ or $\overline Q$  normally
fragments independently;  however, there is a finite probability that
it will combine with a spectator valence quark in the final state to
produce a leading hadron. Coalescence is expected to dominate when the
valence quark and the produced heavy quark have the same velocity. The
coalescence amplitude should be largest at small relative rapidity
since the invariant mass of the $\overline Q q $ system is minimal and
the binding amplitude of the heavy meson wavefunction is maximal. This
picture of coalescence is also consistent with ``heavy quark symmetry"
\cite{isw,cale}. A similar final-state coalescence mechanism is
contained in PYTHIA, a Monte Carlo program based on the Lund string
fragmentation model\cite{PYT}.  Its string mechanism produces some
charmed hadrons with a substantially larger longitudinal momentum than
the charmed quarks originally produced by the fusion processes. At
large $x_f$ and low invariant string masses, the produced $D^-$ or
$D^0$ inherits all the remaining projectile momentum while $D^+$,
$\overline D^0$ production is forbidden.  However, PYTHIA substantially
overestimates the observed asymmetry, ${\cal A}(x_f),$  particularly at
low $x_f$. It also results in ${\cal A}(p_T^2) \sim 0.3$ for
$0<p_T^2<10$ GeV$^2$, overestimating the effect seen in the
$x_f$-integrated data.  This larger asymmetry produced by PYTHIA is due
to a general excess of $D^-$ compared to $D^+$ production, presumably
arising from differences in $c$ and $\overline c$ quark fragmentation
and is not a general feature of final-state coalescence models.
 
In the usual picture of leading charm hadroproduction and in PYTHIA, it
is implicitly assumed that coalescence is strictly a final-state
phenomenon. In fact, the coalescence of the charm quark and a
projectile valence quark may also occur in the initial state.  For
example, the $\pi^-$ can fluctuate into a $| \overline u d c \overline
c \rangle$ Fock state.  The most important fluctuations occur at
minimum invariant mass ${\cal M}$ where all the partons have
approximately the same velocity.   Characteristically, most of the
momentum is carried by the heavy quark constituents of these Fock
states. As viewed from the target rest frame, the intrinsic charm
configurations can have very long  life-times, of order $\tau = 2
P_{\rm lab}/ {\cal M}^2$ where $P_{\rm lab}$ is the projectile
momentum. Intrinsic charm hadroproduction occurs dominantly when the
spectator quarks interact strongly in the target\cite{BHMT}, explaining
why large $x_f$ charm production on nuclear targets is observed to have
a strong nuclear dependence, similar to that of the inelastic
hadron-nucleus cross section. Since the charm and valence quarks have
the same rapidity in an intrinsic charm Fock state, it is easy for them
to coalesce into charmed hadrons and produce leading particle
correlations at large $x_f$ where this mechanism can dominate the
production rate.
 
The leading charm asymmetry must be a higher-twist effect or it would
violate PQCD factorization.  Final-state coalescence is higher twist
since only a small fraction of the fusion-produced heavy quarks will
combine with the valence quarks.  Intrinsic heavy quark production is
also higher twist because the virtual configurations in the projectile
wavefunction must be resolved during their limited lifetime.  The cross
section decreases with extra powers of $1/m_Q$ relative to
leading-twist fusion.  From a general quantum-mechanical standpoint,
both types of higher-twist mechanisms, coalescence of fusion-produced
charm in the final state and coalescence of the intrinsic charm
configurations in the initial state, must occur in QCD at some level.
Recently Vogt and I\cite{ref16} have computed the asymmetry within a
two-component model: parton fusion with coalescence, and intrinsic
charm with valence-quark recombination\cite{VBH2}. In this model the
coalescence mechanism is treated simply as a probabilistic process
where the momenta simply add to form the charmed hadron.  More recently
Quack, Vogt and I\cite{BQV} have treated coalescence at the amplitude
level as a quantum mechanical process.
 
The effects of coalescence can also explain the qualitative features of
$J/\psi$ suppression seen in heavy ion\cite{NA38} and $p A$
collisions\cite{E772}.  When charm quarks are produced they will often
coalesce with comoving partons to produce open charm hadrons, thus
reducing the probability  of quarkonium
production\cite{BrodskyMueller}. The magnitude of the suppression of
quarkonium production is enhanced in the nuclear fragmentation regions
or in events with high transverse energy  when the nuclear target
produces a high density of co-movers. Vogt and I\cite{ref16} have shown
that this effect can explain the main features of $J/\psi$ and
$\Upsilon$ nuclear suppression observed in the NA38 and E772 experiments.
 
\vspace*{1pt}\textlineskip
\section{Anomalous Charm Decays}
\vspace*{-0.5pt}
\noindent
Another unusual feature of charmonium physics is the pattern of
exclusive two body decays of the $J/\psi$ and $\psi^\prime.$  For
example, the branching ratio $BR(J/\psi\to \rho \pi) = 1.28(10)
10^{-2}$ whereas there is only an upper limit for the $\psi^{\prime} $:
$BR(\psi^{\prime} \to \rho \pi) < 8.3\  10^{-5}$ at 90\%\ confidence
level.   Normally one would expect that the branching ratios for any
two-body hadronic channel would proceed through three-gluon
intermediate states. The magnitude is controlled by the charmonium
wavefunction at small $c \bar c$ separation and thus should track with
the lepton pair rates; \ie, the $\psi^{\prime}$ rate should be $15\%$
that of the $J/\psi.$ For most decays this appears to be true; however,
hadron helicity conservation at the perturbative QCD level predicts
that vector-pseudoscalar channels  should be strongly
suppressed\cite{BLT}. Thus the suppression of $\psi^\prime \to \rho
\pi$ is expected, but the large rate for $J/\psi \rightarrow \rho\pi$
is truly anomalous. A good account of this physics and a review of the
latest data from BES\cite{BES} is given by Olsen in these
proceedings\cite{Olsen}.
 
The simplest explanation for the large $J/\psi \to \rho \pi$ decay rate
is the postulate that the $J/\psi$ is mixed with a nearby glueball
state with the same $1^{--}$ quantum numbers. Such a glueball, called the
Omicron, ${\cal O}$, could decay preferentially to vector-pseudoscalar
channels\cite{HouSoni}. This explanation\cite{HouSoni,BLT} also would
imply that an increase in $e^+ e^-  \to \rho \pi$  should be observable
over background at the  mass and width of the $\cal O$. Thus far the
sensitivity of the experiments has not been sufficient to detect the
presence the ${\cal O}$ if its width is greater than $10$
MeV\cite{Olsen}. The possible effect of color-octet multigluon Fock
states in the charmonium wavefunction on its decay  also needs further
exploration.

\vspace*{1pt}\textlineskip
\section{Channels and Its Production at Large $p_T$}
\vspace*{-0.5pt}
\noindent
As I have discussed in this brief review, there are a remarkable number
of anomalies associated with the production and decay of heavy quark
systems.  A common thread in the proposed explanations for the
anomalies within QCD are non-perturbative and higher Fock state effects
which can cause mixing of hadronic states and enhanced interactions
near threshold and at low relative velocities. There are other
important consequences which require experimental verification,
including nuclear-bound quarkonia, intrinsic heavy quark contributions
to  structure functions and target fragmentation, anomalous spin
correlations, enhanced rates near heavy quark thresholds, leading
particle effects, and unusual nuclear target dependencies.

\vspace*{1pt}\textlineskip
\section{ACKNOWLEDGEMENTS}
\vspace*{-0.5pt}
\noindent
I wish to thank Howard Goldberg, Tom Imbo, and Wai-Yee Keung for
organizing this interesting meeting. I also thank M. Beneke, M. Gill,
P. Hoyer, G. P. Lepage, B.-Q. Ma, A.~Mueller, F. Navarra, E. Quack, H.
Quinn, and  R. Vogt for helpful conversations. This work was supported
by the Department of Energy under contract \nopagebreak number
DE--AC03--76SF00515.

\pagebreak 
\vspace*{1pt}\textlineskip

\end{document}